\RequirePackage{fix-cm}
\documentclass[natbib, smallextended]{svjour3a}       
\smartqed  
\usepackage{graphicx, pdflscape, hhline, subfiles}
\usepackage{amsmath}
\usepackage{float, multirow, tabu}
\usepackage{amsfonts}
\usepackage{amssymb}

\newcommand{\vv}[1]{#1} 

\usepackage[]{lineno}
\journalname{Climatic Change}
\begin{document}
\title{Improved Representation of Ocean Heat Content in Energy Balance Models}
\author{B.T. Nadiga and N.M. Urban}
\institute{B.T. Nadiga \at
Los Alamos National Lab, Los Alamos, NM 87544\\
Tel.: 5056679466\\
\email{balu@lanl.gov}
\and
N.M. Urban \at
Los Alamos National Lab, Los Alamos, NM 87544
}
\date{}
\maketitle
\begin{abstract}
  Anomaly-diffusing energy balance models (AD-EBM) are routinely   employed to analyze and emulate the warming response of both   observed and simulated Earth systems. We demonstrate a deficiency in   common multi-layer as well as continuous-diffusion AD-EBM variants:   They are unable to, simultaneously, properly represent surface warming and the vertical distribution of heat uptake. We show that this inability is   due to the diffusion approximation.  On the other hand, it is well   understood that transport of water from the surface mixed layer into   the ocean interior is achieved, in large part, by the process of   ventilation---a process associated with outcropping isopycnals. We,   therefore, start from a configuration of outcropping isopycnals and   demonstrate how an AD-EBM can be modified to include the effect of   ventilation on ocean uptake of anomalous radiative forcing. The   resulting EBM is able to successfully represent both surface warming   and the vertical distribution of heat uptake, and indeed, a simple   four layer model suffices. The simplicity of the models   notwithstanding, the analysis presented and the necessity of the   modification highlight the role played by processes related to the   down-welling branch of global ocean circulation in shaping the   vertical distribution of ocean heat uptake. 
\end{abstract}

\section{Introduction}

Energy balance models (EBM) have long been used in climate science as
conceptual tools to understand heat transfer within the Earth system
\citep[e.g.,][]{budyko1969effect,sellers1969global,hoffert1980role,
  schneider1981atmospheric,north1981energy, murphy1995transient,
  gregory2000vertical}, as computationally-efficient emulators of and
diagnostic tools for more complex Earth system models\footnote{e.g.,
  because of their computational intensity, ESMs themselves are never
  run long enough to directly assess climate sensitivity.} (ESMs)
\citep{wigley1992lmplications,raper1996emulation,houghton1997introduction,
  raper2001use,meinshausen2011emulating,geoffroy2012quantifying}, and
to explore and quantify uncertainties in climate change projections
\citep{urban2010probabilistic,padilla2011probabilistic,aldrin2012bayesian,johansson2015equilibrium,bodman2016bayesian}.
EBMs represent changes in climate as the transfer of heat between
appropriately aggregated units, and where the units are loosely
referred to as layers when they are distributed in the vertical
dimension alone (cf. isopycnal layers) and as boxes if other
dimensions are involved as well.  As an example of the former, in
two-layer models that are widely used \citep[e.g., see][and references
therein]{schneider1981atmospheric, gregory2000vertical,
  held2010probing} both for climate projection purposes and to emulate
Atmosphere-Ocean General Circulation Models (AOGCM), the atmosphere,
the land surface, and the oceanic mixed layer components of the earth
system are aggregated into one surface layer with a smaller thermal
inertia and it interacts with a second layer that is representative of
the sub-surface ocean and which has a larger thermal inertia.  On the
other hand, EBMs may allow for a coninuous representation in the
relevant dimension \citep[e.g.,][and others]{budyko1969effect,
  sellers1969global, raper2001use, meinshausen2011emulating}. Thus, in
the context of upwelling-diffusion or pure-diffusion in the vertical,
such EBMs consist of many layers (typically $\sim$40) with the intent
of representing the limit of a continuous vertical dimension.  Further
combinations are easily realized.

In its use as a physics-based emulator, we would ask that an EBM be
able to reproduce both ESM-simulated and observed surface {\em and
  subsurface warming} in a range of scenarios.  Diffusion-class models
have shown skill at reproducing observed surface warming and overall
ocean warming as well as warming behaviors of ESMs under historical
and future CO$_2$ emissions scenarios \citep[e.g., see][and
others]{gregory2000vertical, raper2001use, raper2002role}.  In
particular, a popular form of a two-layer model has proven to be
capable of capturing the typical two-timescale surface temperature
response seen in abruptly forced AOGCM simulations \citep[e.g., see][and
others]{raper2001use, friend2011response, geoffroy2013transient}
across a range of AOGCMs.  However, how well that model can represent
the vertical structure of ocean heat uptake has not received much
attention. We find that the two-layer model is unable to
  simultaneously properly represent surface warming responses and
  observational estimates of 0-700m and 700-2000m heat
  storage. Perhaps more surprising is that we find that more highly-resolved
  diffusion-class models are similarly deficient.\footnote{In order to exclude the possibility that our
    results are due to poor parameter tuning in the EBMs we consider,
    we apply a Bayesian statistical framework and Monte Carlo sampling
    to widely explore the space of parameter uncertainties.}

On the other hand, from the point of view of dynamical processes
underlying global ocean circulation, this last finding---that
vertically-resolved EBMs are unable to simultaneously fit both the
surface response and the vertical distribution of ocean heat
uptake---is not surprising for the following reasons: (a) From the point
of view of ocean dynamics, it is well understood that
transport of water from the surface mixed layer into the ocean
interior is achieved, in large part, by the process of ventilation---a
process associated with outcropping isopycnals. That is, when
isopycnals outcrop at high latitudes, water masses that are at the
surface at high latitudes are able to move laterally back to middle
and low latitudes at deeper levels \citep[e.g., see Chapter 4
of][]{pedlosky2013ocean}. (b) The deep ocean loses heat to the atmosphere at high
latitudes through convective instabilities \citep[e.g., see Chapter 7
of][]{talley2011descriptive}, and when the stability of
the water column is incrementally enhanced in a warming scenario (that
is, the water column continues to remain unstable, but its degree of
instability is slightly reduced), the heat loss from the deep ocean is
slightly reduced leading effectively to a warming of the deep. 
Indeed, the effects of such processes have been verified in
various climate models \citep[e.g., see][and
others]{gregory2000vertical, raper2001use,
  raper2002role}.

Clearly, while there are many directions in which one could improve
the realism of an EBM's warming response, in this work we present
one simple modification that enables simultaneously fitting a wide
range of scenarios: Outcropping of isopycnals/neutral surfaces is a
common feature of global ocean circulation \citep[e.g.,
see][]{talley2011descriptive, pedlosky2013ocean}. Nevertheless, this
configuration has not received much attention in the context of
EBMs. Indeed, we find that a cure to the problem identified in AD-EBMs
lies in considering such a configuration. We, therefore, start from a
conceptual configuration of outcropping isopycnals. Next, we perform a
transformation of coordinates in order to cast this configuration into
a framework of horizontally-averaged layers, a framework that is more
commonly encountered in the EBM context. We then show mathematically
that when such a configuration of outcropping isopycnal/neutral
surfaces is viewed in terms of horizontally-averaged layers, non-local
interactions between near-surface and deeper ocean layers result. We
then show that such a modification, a modification that represents the
lowest order effects of outcropping isopycnal/neutral surfaces, is
sufficient to simultaneously capture surface warming and deeper ocean
heat uptake.  Indeed, we find that it is not necessary to use a fully
one-dimensional model; a four-layer model suffices.


\begin{figure}[h!]
\includegraphics[width=\textwidth]{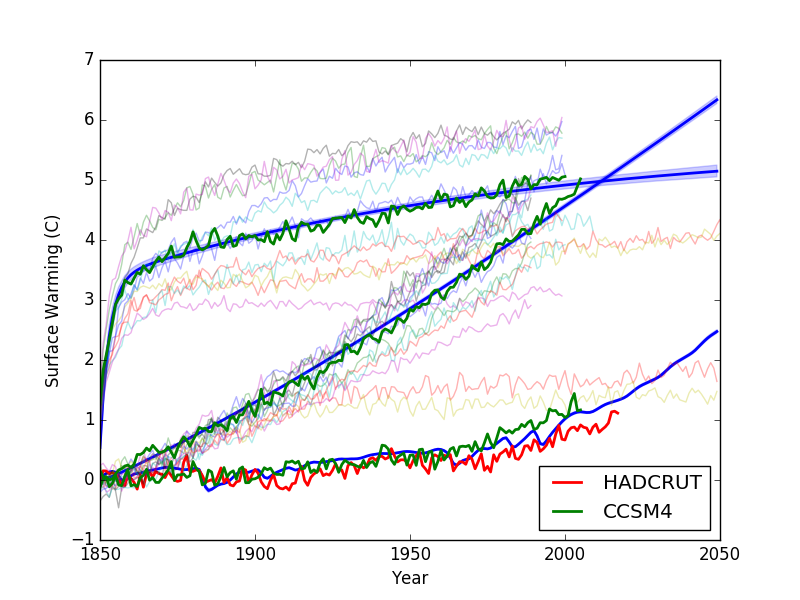}
\caption{Here, the
  popular two layer EBM given by Eqs.~\ref{eq-upper} and
  \ref{eq-lower} is calibrated against globally-averaged and
  annually-averaged SAT response of the CCSM4 model simulation of the
  abrupt4x CO$_2$ experiment of the CMIP5 protocol using a Bayesian
  framework. Mean and 95 percentile envelopes of the posterior predictive distribution of
  SAT using the calibrated SCM (blue) is compared against the SAT evolution
  in the full CCSM4 simulations for the 1\% CO$_2$ and historical
  forcing experiments (green). The SAT response of a handful of other
  AOGCMs in the abrupt4x and 1\% CO$_2$ forcing CMIP5 experiments are shown in
  lighter lines.}
\label{2lyr-sat}
\end{figure}

\section{Methodology and Results}
\subsection{The Two-Layer EBM}
A particular form of simple climate model (SCM) that has been popular in summarizing
integral thermal properties of 
AOGCMs and ESMs is the 
anomaly-diffusing energy-balance model (AD-EBM)
(e.g, see \citet{schneider1981atmospheric, gregory2000vertical, held2010probing}).  In this
horizontally-integrated model of the climate system, the upper (surface)
layer is comprised of the oceanic mixed layer, the atmosphere and the land
surface and the bottom layer represents the ocean beneath the
mixed layer.  Evolution of the upper surface heat content anomaly per
unit area is given by
\begin{linenomath*}
\begin{align}
C_u\frac{dT_u}{dt} &= {\cal F} - \lambda T_u - \gamma\left(T_u-T_d\right) 
\label{eq-upper}
\end{align}
\end{linenomath*}
where ${\cal F}$ represents the effective (anomalous) radiative
forcing (e.g., as due to green-house gases recently), $\lambda$ represents the climate
feedback parameter (e.g., see \citet{knutti2008equilibrium}). Exchange
of heat between the surface layer and the ocean beneath is
parameterized by the difference in temperature between the two layers,
with $\gamma$ being the constant of proportionality.
Similarly, evolution of the subsurface ocean heat content (OHC) anomaly
(again per unit area) is given by
\begin{linenomath*}
\begin{align}
C_d\frac{dT_d}{dt} &= \gamma\left(T_u-T_d\right).
\label{eq-lower}
\end{align}
\end{linenomath*}
The heat capacities $C_u$ and $C_d$ can be equivalently expressed in
terms of ocean layer depths, $h_u$ and $h_d$ as $C_{(\cdot)} = c h_{(\cdot)}$,
where $c$ is the volumetric heat capacity with a nominal value of
4.1$\times$ 10$^6$ Jm$^{-3}$K$^{-1}$.

\subsection{Skillful Representation of Surface Warming in the Two-Layer EBM}
This two-layer model has been used extensively for estimating climate
sensitivity of AOGCMs and ESMs (e.g., see
\citet{held2010probing,geoffroy2012quantifying}, and
others). Figure~\ref{2lyr-sat} shows an example of its typical use. In
this figure, we have calibrated the four EBM parameters ($\lambda$,
$\gamma$, $h_u$, $h_d$) using the Community Climate System Model v. 4.0
(CCSM4) AOGCM's annual global surface air temperature (SAT) output from
an abrupt CO$_2$ quadrupling (abrupt4x) experiment\footnote{We note
  that the intention of the CMIP abrupt 4$\times$CO$_2$ forcing
  experiment is one of calibration and diagnosis of equilibrium
  climate sensitivity since post-industrial surface warming itself is
  compatible with a wide range of climate sensitivities} as the {\em
  sole} calibration data.  The
parameters are inferred probabilistically from the CCSM4 SAT time
series using Bayesian Monte Carlo sampling, assuming a first-order
autoregressive noise model to capture the AOGCM natural variability not
represented in the EBM (e.g., as in \citet{urban2010probabilistic} or
\citet{nadiga16}; see online supplement section titled ``Further
Methodological Details'' for further details).  Bayesian inference
allows us to explore the full set of EBM parameters that are capable
of fitting the AOGCM SAT data, to ensure that our results are not
fragile with respect to one particular choice of parameters.

Next, we validate the calibration of the two-layer model by
considering new forcing scenarios---1\%/year CO$_2$ increase and
historical forcings.  By propagating the, calibrated, joint posterior
distribution of parameters, we produce posterior predictive
distributions of the respective SAT responses.  The SAT responses and
their 95 percentile ranges are shown in Fig.~\ref{2lyr-sat} (see
Sec.~A of ES for further details).  The good correspondence between
the posterior predictive distribution of SAT in the 1\% CO$_2$ (RMS
error of 0.11 C) and historical forcing experiments (RMS error of 0.16
C), as compared to the corresponding AOGCM results attest to both the
reasonableness and the usefulness of the model. The SAT response of a
handful of other AOGCMs in the abrupt4x and 1\% CO$_2$ forcing CMIP5
(Coupled Model Intercomparison Project Phase 5;
\cite{taylor2012overview}) experiments (lighter lines) and the Hadley
center estimate (HadCRUT4) are shown for future reference.

\subsection{Ocean Heat Uptake Considerations}
Use of SCMs like in the previous example is well established. However
since in excess of 90\% of the anomalous radiative forcing is
sequestered in the world oceans \citep[e.g.,][and
others]{domingues2008improved, levitus2012world,
  balmaseda2013distinctive} and since the nature of ocean heat uptake
is important in determing surface response, we next examine the
ability of the two-layer EBM to represent aspects of the ocean heat
uptake. For this, the main data product we consider is the
sequestration of heat in the top 700 m and between 700 m and 2 km in
the global ocean, as provided by the National Oceanic Data Center (NODC)
\citep{levitus2012world}. We compare these two observation-based
timeseries against the corresponding timeseries in the SCMs.

We have considered a specific AOGCM, CCSM4, for our example here and
it's clear that we could have chosen any other AOGCM from the CMIP5 model
ensemble. While the particular choice of the AOGCM whose response we use
to calibrate the SCM will affect details of the results presented, we
believe that the importance of this choice is secondary.
Figure~\ref{2lyr-sat} shows in light lines the spread of SAT in the
abrupt4x and 1\% CO$_2$ CMIP5 experiments from a handful of other
modeling centers as well. While there is considerable spread in the
response of the different models in these experiments, the spread in
the averaged response of the different models in the historical
forcing simulations is much less given the required tuning of AOGCMs to
best match the current state of the Earth's climate system
\citep[e.g.,][]{mauritsen2012tuning}. Multi-model AOGCM differences are
therefore not likely to account for the ability of an SCM, fit to a
particular AOGCM, to reproduce historical climate trends.

\begin{figure}[h!]
\includegraphics[width=\textwidth]{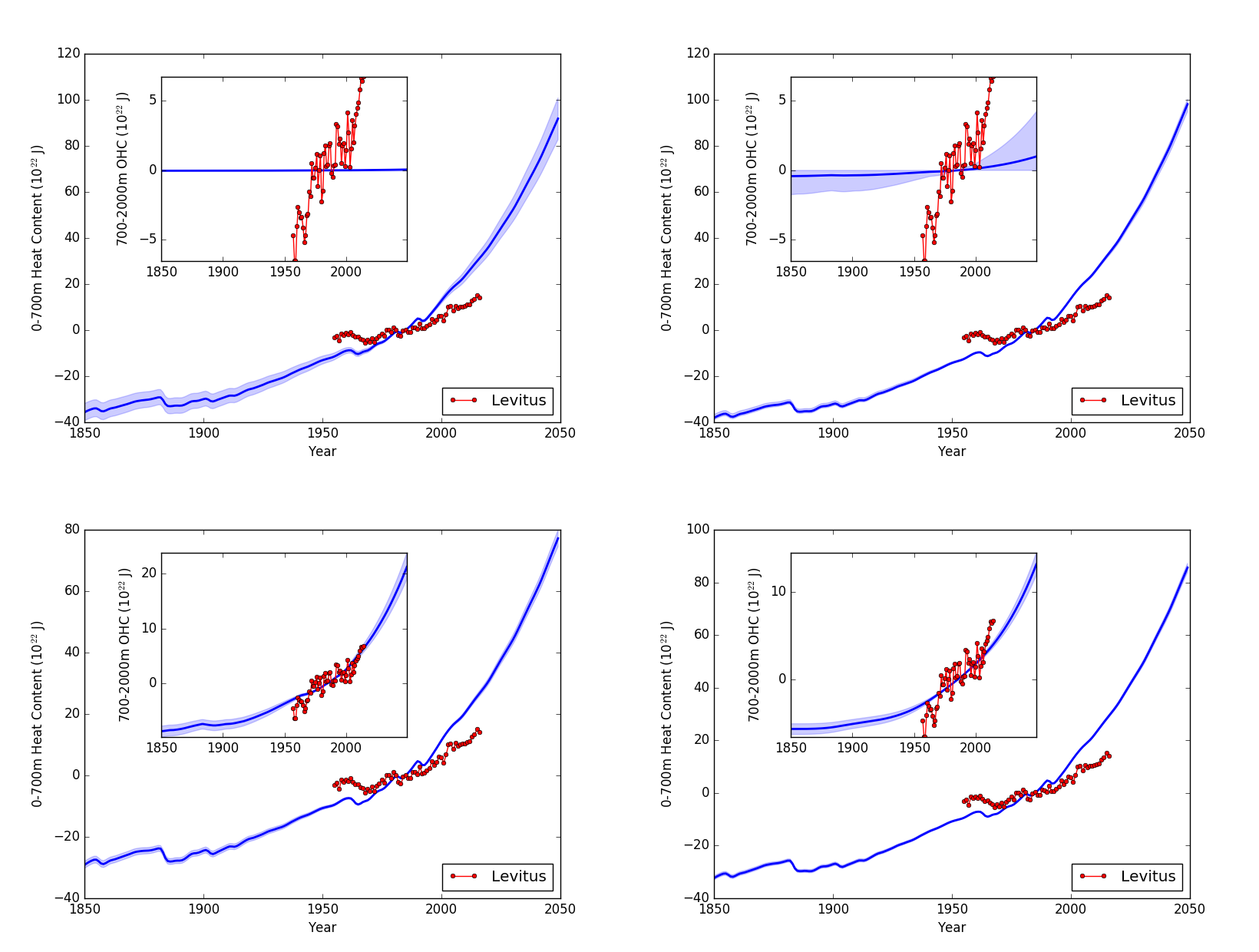}
\caption{Comparison of the SCM representation of OHC against the
  observational estimate of 0-700m and 700-2000m OHC
  \citep{levitus2012world}. Top-left: two-layer EBM calibrated against
  CCSM4's SAT response in the abrupt4x CMIP5 experiment. Top-right:
  calibrated against abrupt4x SAT and RN. Bottom-left: calibrated
  against abrupt4x SAT and RN and the observational estimates of OHC
  on using the historical forcing . Bottom-right: In a 60 layer model
  with calibration as in bottom-left.}
\label{ohc1}
\end{figure}

In the top-left panel of Fig.~\ref{ohc1}, we compare the 0-700m (in
the main plot) and 700-2000 m OHC (in the inset
plot) in the previous experiment\footnote{All experiments are run with
  each of the three forcings under consideration (abrupt4x, 1\%, and
  historical), and the quantities of interest (QoI) are the SAT with
  each forcing and the 0-700m and 700-2000m OHC with historical
  forcing. While we also considered RN, those results are left out for
  brevity.}  against the NODC \citep{levitus2012world} observational
estimates; again the posterior mean and 95\% envelopes are
shown.\footnote{In the two-layer EBM, warming of the ocean extends
  only to the depth of the two layers. That is, there is no warming
  below the depth of the second layer. Furthermore, the depth of the
  top layer should be thought of as roughly equivalent to the
  globally-averaged mixed-layer depth (O(100m); $\ll 700m$) The 0-700
  m and 700-2000 m warming is obtained using linear interpolation.}
The SCM estimates are seen to fit the observational estimates very
poorly.

This may not be surprising considering that in the previous experiment
there was no constraint on heat storage in the climate system; as
shown in Table. 1, calibration data comprised of only the surface
warming in the abrupt4x AOGCM experiment. After all, the main motivation
and the typical use of the two-layer EBM has been to capture the two
timescales that are thought to characterize the surface warming
response (e.g, see \citet{gregory2000vertical} or
\citet{held2010probing}).  

\subsection{Constraining the EBM's Sub-surface Heat Content}
Summing Eqs.~\ref{eq-upper} and \ref{eq-lower} results in 
\begin{linenomath*}
\begin{align}
\frac{d{\cal H}}{dt} = {\cal F} - \lambda T_u;\quad 
{\cal H} = c (h_uT_u + h_dT_d)
\label{eq-hc}
\end{align}
\end{linenomath*}
and Eq.~\ref{eq-hc} represents the evolution of the
heat content anomaly per unit area.
The right hand side of Eq.~\ref{eq-hc} is the radiative
imbalance (RI) or equivalently radiative non-equilibrium (RN). 

Given the poor fit of OHC when the two-layer EBM is calibrated using
the SAT response of the EBM in the abrupt4x experiment, we repeat the
previous experiment, but now calibrating against both the SAT and
radiative imbalance responses of the AOGCM. (A list of the experiments
considered here is shown in Table 1; a few other supporting
experiments discussed in the online supplement are shown in a
table there.) While the SAT behavior is similar to that in
Fig.~\ref{2lyr-sat} (not shown, but see Sec.~2 of supplement for figure and
details), the OHC comparison is shown in the top-right panel of
Fig.~\ref{ohc1}. It is seen that constraining the total heat storage
leads to only minor improvements in the two layer EBM's representation
of ocean heat uptake.  However, from the point of view of calibration
itself, the differences are more dramatic: Not only are the EBM
parameters inferred to be significantly different in the two
experiments, but the inferences themselves are sharper and
correlations that exist between model parameters in experiment 1 are
seen to be greatly reduced in experiment 2 (see Sec.~2 of supplement for
details).

\subsection{A Shortcoming of the Two-Layer EBM}
Since constraining overall heat storage in the abrupt4x CO$_2$ forcing
experiment was seen to be insufficient to
elicit an observationally-consistent heat storage response in the
historical forcing experiment, we consider experiment 3, wherein we
calibrate the two layer EBM against the AOGCM's SAT and RN responses in the
abrupt4x CO$_2$ forcing experiment, and calibrate against the
observational estimate (0-700m and 700-2000m) of OHC itself, when the
SCM is forced with an estimate of the historical forcing followed by
the RCP8.5 scenario \citep{meinshausen2011rcp} to 2050.

In experiment 3, the simple two layer EBM's representation of OHC
(bottom left panel of Fig.~\ref{ohc1}) is seen to be better with
improvements in the representation of the 700-2000m heat
content. However, the 0-700m warming continues to be poorly
represented. Attendant with the partial improvement in the
representation of heat storage, unsurprisingly, is a slight
degradation in the ability of the EBM to fit the SAT response in the
abrupt4x CO$_2$ experiment (see Sec.~2 of supplement for further
details and discussion).

We consider the inability of the two layer model to simultaneously
properly represent the surface warming and the 0-700m and 700-2000m
heat storage a shortcoming of the model and seek a simple extension of
the model that will not suffer from this shortcoming. For brevity,
when we refer to {\em the} problem or {\em the} shortcoming in the
rest of the article, we are specifically referring to the shortcoming
just described. 
 
\subsection{Persistence of the Shortcoming in Vertically-Resolved AD-EBMs}
To investigate the possibility that it is the lack of resolution in
the vertical that is the cause of the inability of the two layer EBM
to represent heat storage in a manner that is consistent with
observations, we considered a three layer extension and a four layer
extension (experiment 4) of the two layer model. In each of these
cases, calibration proceeded as in experiment 3.  The close similarity
in the SAT and heat storage responses of the three (not shown) and
four layer models (see Table 1) to that in the two-layer suggests that
a lack of resolution in the vertical may not be the {\em direct} cause
of the problem.

\begin{landscape}
\begin{table}
\caption{A brief overview of the experiments conducted. Units for RMS error of SAT is degrees C, and units for RMS error of OHC is $10^{22}$ J.}
\label{tab1}
\centering
{\tabulinesep=1.2mm
\begin{tabu}{|c|c|c|c|c|c|c|c|c|c|c|}
\hline
\multirow{2}{*}{Expt. \#} & \multirow{2}{*}{SCM}     & \multicolumn{3}{|c|}{Calibration Details} &  \multicolumn{4}{c|}{RMS Error: SAT}&  \multicolumn{2}{|c|}{OHC}\\
\cline{3-11}
                          &                          & Forcing                                   & AOGCM QoI & Obs. Data     & 4x&1\%&HIS&OBS&$\le$700&$>$700\\
\hhline{|=|=|=|=|=|=|=|=|=|=|=|}
1                         & 2L                       & 4x                                        & SAT          &               &0.13&0.11&0.16&0.18&7.5&3.2\\
\hline
\multirow{1}{*}{2}        & 2L                       & \multirow{1}{*}{4x}                       & SAT, RN      &               &0.14&0.12&0.16&0.18&8.5&3.1\\
\hline
\multirow{3}{*}{3}        & \multirow{3}{*}{2L}      & 4x                                        & SAT, RN      &               & 0.21&0.14&0.16&0.18&5.8&1.5     \\
\cline{3-11}
                          &                          & \multirow{2}{*}{Historical}               &              & 0-700m OHC    &  \multicolumn{6}{c|}{Degraded 4x SAT}\\
\cline{5-5}
                          &                          &                                           &              & 700-2000m OHC &  \multicolumn{6}{c|}{Improved but Poor OHC}\\
\hline
\multirow{3}{*}{4}        & \multirow{3}{*}{4L}   & 4x                                        & SAT, RN      &               & 0.17&0.13&0.16&0.18&5.7&1.4\\
\cline{3-11}
                          &                          & \multirow{2}{*}{Historical}               &              & 0-700m OHC    & \multicolumn{6}{c|}{Reasonable SAT}\\
\cline{5-5}
                          &                          &                                           &              & 700-2000m OHC & \multicolumn{6}{c|}{Poor OHC}\\
\hline
\multirow{3}{*}{5}        & 60L & 4x                                        & SAT, RN      &               &0.18&0.13&0.16&0.18&6.5&1.7\\
\cline{2-11}
                          &  \multirow{2}{*}{(many variants)}                        & \multirow{2}{*}{Historical}               &              & 0-700m OHC    &\multicolumn{6}{c|}{Reasonable SAT}\\
\cline{5-5}
                          &             &                                           &              & 700-2000m OHC & \multicolumn{6}{c|}{Poor OHC}\\
\hline
\multirow{3}{*}{6}        & \multirow{3}{*}{4LE}     & 4x                                        & SAT, RN      &               &0.18&0.13&0.16&0.18&1.4&1.5\\
\cline{3-11}
                          &                          & \multirow{2}{*}{Historical}               &              & 0-700m OHC    &\multicolumn{6}{c|}{Reasonable SAT}\\
\cline{5-5}
                          &                          &                                           &              & 700-2000m OHC &\multicolumn{6}{c|}{Reasonable OHC}\\
\hline

\end{tabu}}
\end{table}
\end{landscape}
Nevertheless, to probe the issue of vertical resolution further, we
considered a number of variants of the continuously stratified version
of the AD-EBM. A few of these experiments are detailed in the
supplement under the section titled ``Experiments with Continuous
AD-EBM Variants''.  \citep{bryan1979water, nadiga16}. In these
experiments, we typically discretized the subsurface thermocline
region down to 2500 m into a series of layers of equal thickness since
observational estimates of warming below 2000 m in the global ocean
tend to be small \citep[e.g.,][]{cheng2017improved}: $<$10\% of the
full ocean warming or $\lesssim$ 3$\times$ 10$^{22}$ J currently.  In
these experiments, the main variations were in the form of vertical or
diapycnal diffusivity assumed, and as to whether the coefficient(s) of
diffusivity itself was (themselves were) fixed at realistic values or
inferred in the calibration procedure. Other variations included
fixing or inferring the depth of the upper layer and allowing for a
(depth-independent) upwelling velocity or not (see supplement for
details and results).\footnote{When a number of the variations mentioned are
  considered in combination, and with a large number of layers, calibration
  of the resulting model can sometimes lead to, e.g., the top layer
  depth being different from the a priori estimate of around 70m. In
  such cases, we performed a companion experiment in which the top
  layer depth is fixed at 70m, and so on.}  Like in the other panels of
Fig.~\ref{ohc1}, the bottom-right panel shows the OHC comparison for
one such model that fit AOGCM responses and observational estimates of
OHC best (Experiment 5g in supplement).

The improvements in going from a two-layer model to a large number of
layers (here 60) is seen to be rather small. The inability of the
continuously stratified AD-EBM may be understood as follows: In the
suite of experiments conducted with the continuous model (and
described in Sec.~3 of
supplement), in the cases which fit AOGCM responses and observational estimates
of OHC best, the inferred value of thermocline diffusivity ranged
between 0.7 and 0.9 cm$^2$/s. These values are much higher than the
accepted levels of 0.1 cm$^2$/s. One reason for these large values of
vertical diffusivity is that the model is attempting to get sufficient
heat down to the 700-2000 m depths in order to satisfy the
observational constraint. And in trying to do so, it is clear from
these experiments that this is causing too large a heating in the
0-700 m depth range---a feature that is consistent across a wide suite
of experiments (described in Sec.~3 of supplement).  That is, assuming that
vertical diffusivity is not varying with time (and there is no reason
to expect it to have varied substantially with time in the post
industrial period) getting observed levels of warming into the
700-2000 m layer while limiting the warming of the 0-700m layer to
observed levels is difficult to achieve in AD-EBMs.

\subsection{Moving Beyond Diffusion of Anomalies: Outcropping Surfaces} 
The inability of the continous variants of AD-EBM model to reasonably
represent the vertical distribution of OHC leads us to next consider
if it is the anomaly diffusion aspect of these models that prevents
the SCM from being able to simultaneously properly represent surface
warming and the 0-700m and 700-2000m heat storage. After all, the
global ocean circulation tends to be highly dynamic in the sense that
it is largely adiabatic outside of the surface layer and bottom
boundary layers, and an anomaly diffusing model attempts to
parameterize the horizontally-averaged effect of the myriad dynamical
processes through diffusion across adjoining layers. This may be
restrictive in the sense that the effect of processes such as
deep-water formation, and subduction and mode water formation
processes on ocean heat storage, when averaged horizontally, may no
longer appear as local diffusion operators. For example, when water in
the midlatitudes whose properties are set by its last interaction with
the atmosphere slides down an appropriate isopycnal, in a horizontally
averaged sense this process will appear as if two non-adjoining layers
are interacting.

\begin{figure}[h!]
\includegraphics[width=\textwidth]{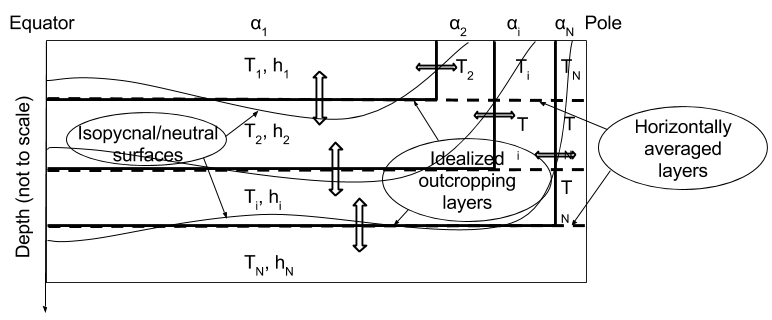}
\caption{Model schematic: In the meridional-depth plane (not to scale)
outcropping isopycnal/neutral surfaces are show in thin continuous
curved lines. These are then idealized into horizontal and vertical
surfaces shown in thick continuous straight lines. Finally the
horizontally-averaged layers are shown in dashed straight lines. We
demonstrate mathematically that purely local interactions in the
isopycnal/neutral coordinates lead to non-local interactions when
horizontally-averaged layers are considered.}
\label{nonlocal-vert}
\end{figure}

To demonstrate this effect, consider a schematic of the meridional-depth cross-section
of the world oceans, as shown in Fig.~\ref{nonlocal-vert}. In this
figure, the outcropping isopycnal/neutral surfaces in thin continuous
lines are idealized into the horizontal and vertical surfaces shown in thick
continuous lines. In Fig.~\ref{nonlocal-vert}, $\alpha_i$ is the
meridional fraction of layer $i$ that outcrops and is exposed to the
atmosphere. The EBM for such layers is:
\begin{linenomath*}
\begin{align}
c\left(\alpha_i\sum_{j=1}^{i-1} h_j +
  h_i\sum_{j=1}^{i}\alpha_j\right)\frac{d{T_i}}{dt} =
  \alpha_i\left({\cal F} - \lambda T_i\right) + \sum_{j=\max(1,
  i-1)}^{\min(N, i+1)} \gamma_{ij}(T_j - T_i)
\label{eq-nl1}
\end{align}
\end{linenomath*}
and where $\gamma_{ij}$ ($= \gamma_{ji}$) parameterizes the
interactions between adjoining layers $i$ and $j$ and the summation terms drop to
0 when the starting lower index is greater than the ending upper
index.

With trivial manipulation, Eq.~{\ref{eq-nl1}} can be rewritten in a
compact symbolic form as
\begin{linenomath*}
\begin{align}
\frac{d{\vv T}}{dt} = \vv C + \vv A \vv T
\label{eq-nl2}
\end{align}
\end{linenomath*}
where $\vv T$ is the vector of temperatures $T_i$ in Eq.~{\ref{eq-nl1}},
$\vv C$ is a constant vector related to the anomalous radiative
forcing and $\vv A$ is a tridiagonal matrix that is independent of
$\vv T$. (Note that $\vv A$
  in Eq.~{\ref{eq-nl2}} is not symmetric even though the right hand side
  of Eq.~{\ref{eq-nl1}} is. This is because each row of the matrix form
  of Eq.~{\ref{eq-nl1}} gets divided by a different constant (the
  coefficient of $\frac{d{\vv T}}{dt}$) in Eq.~{\ref{eq-nl1}}.)
Next, consider horizontal averaging of temperature in
Fig.~{\ref{nonlocal-vert}} to go from the idealized outcropping
layers (shown in thick continuous lines) to horizontally-averaged layers (shown
in dashed lines):
\begin{linenomath*}
\begin{align}
\sum_{j=i+1}^{N}\alpha_j T_j +
T_i\sum_{j=1}^{i}\alpha_j
= \widetilde T_i
\label{eq-transform}
\end{align}
\end{linenomath*}
where $\widetilde T_i$ is the temperature of the $i^{th}$
horizontally-averaged layer, and again the same rules hold for
summation. Again, Eq.~{\ref{eq-transform}} can be written in a compact
form as:
\begin{linenomath*}
\begin{align}
\vv B \vv T = \widetilde {\vv T}
\label{eq-transform2}
\end{align}
\end{linenomath*}
where by virtue of the limits of the summation in the first term on
the left hand side of Eq.~{\ref{eq-transform}}, $\vv B$ is seen to be
upper triangular. In the generic case of $\sum_{j=1}^{i}\alpha_j\ne
0$, Eq.~{\ref{eq-transform2}} can be inverted as
\begin{linenomath*}
\begin{align}
\vv T = \vv B^{-1} \widetilde {\vv T}.
\label{eq-transform3}
\end{align}
\end{linenomath*}
$\vv B^{-1}$ is upper triangular because $\vv B$ is upper triangular. 
Inserting Eq.~{\ref{eq-transform3}} in Eq.~{\ref{eq-nl2}} and simplifying leads to
\begin{linenomath*}
\begin{align}
\frac{d \widetilde {\vv T}}{dt} = \widetilde {\vv C} + \widetilde {\vv A} \widetilde{\vv T}.
\label{eq-nl3}
\end{align}
\end{linenomath*}
While $\vv A$ is tridiagonal (see Eq.~{\ref{eq-nl1}}), $\widetilde{\vv
  A}$ is not: To see that $\widetilde{\vv A}$ is not
  tridiagonal, consider that
\begin{linenomath*}
\begin{align}
\widetilde A_{ij} 
= \sum_{k=1}^n \sum_{l=1}^n  B_{ik}A_{kl}B^{-1}_{lj} 
= \sum_{k=i}^n \sum_{l=1}^j  B_{ik}A_{kl}B^{-1}_{lj} 
\label{eq-nl4}
\end{align}
\end{linenomath*}
where the second equality is due to the upper triangular nature of
$\vv B$ and ${\vv B}^{-1}$ ($B_{ij}=B_{ij}^{-1}=0$ for $i>j$).
Therefore if $i > j + 1$, the values of $k$ and $l$ in the above sum
satisfy the inequality $k \ge i > j + 1 \ge l + 1$, meaning that
$k > l + 1$. Therefore $A_{kl} = 0$ for all the terms in the above
sum for $i > j + 1$, making ${\widetilde A}_{ij} = 0$ for $i > j +
1$.

That is while the outcropping isopycnal/neutral layers interact purely
locally in isopycnal/neutral coordinates (tridiagonal $\vv A$ in
Eq.~{\ref{eq-nl2}}), the same dynamic when represented using
horizontally averaged layers leads to nonlocal interactions (e.g., the
top layer interacts with as many of the underlying layers as there are outcropping isopycnals.)

\subsection{Results with the Modified EBM}
Following the mathematical development above, we next evaluate if the
modified EBM given by Eq.~\ref{eq-nl3} is capable of overcoming the
shortcoming identified in the AD-EBMs. As previously mentioned,
$\widetilde{\vv A}$ not being tridiagonal amounts to allowing for
interactions between non-adjoining, horizontally-averaged layers.  As
for adjoining layers, we again parameterize such interactions in
terms of the difference in their temperature anomalies.  The simplest
extension then consists of a three layer model with a new interaction
between the top layer and the deepest layer. Within the range of
experimentation that we conducted, we find that such a three layer
extension is incapable of adequately resolving the shortcoming; for
that we need at least four layers.

\begin{figure}[h!]
\includegraphics[width=\textwidth]{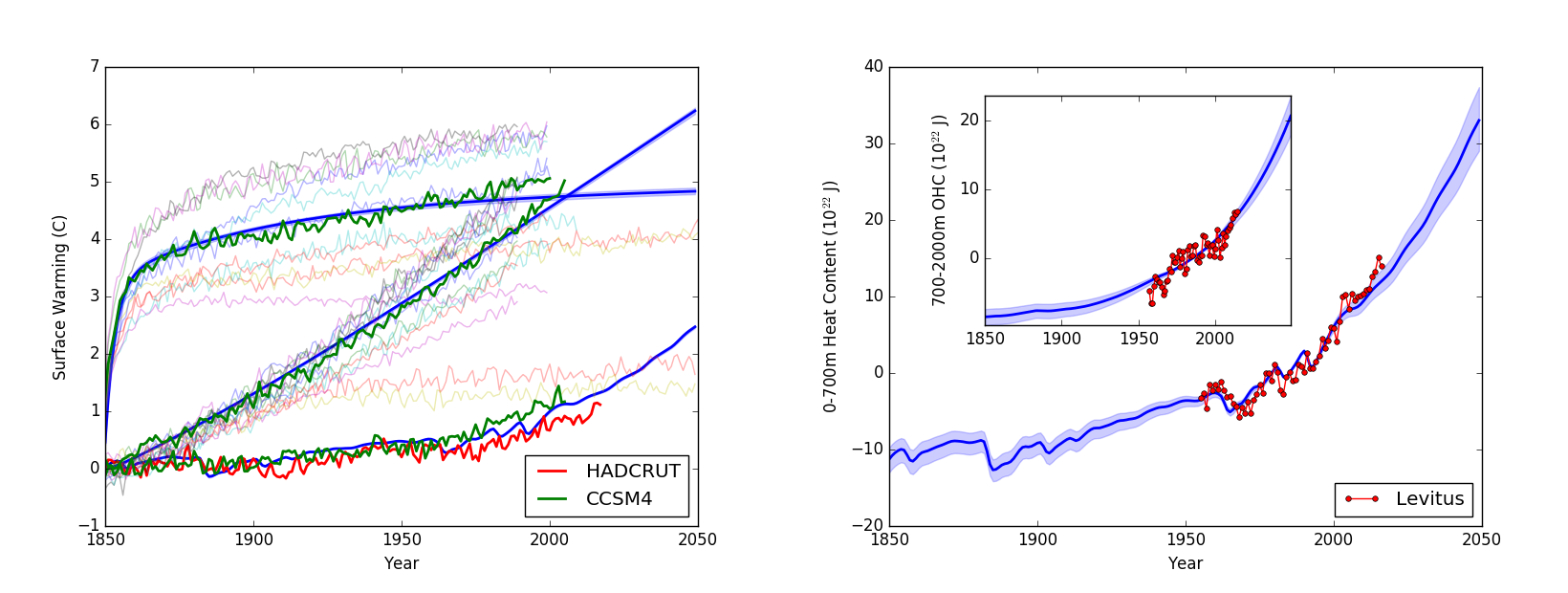}
\caption{In Experiment 6, a four layer model with non-local
  interactions is seen to be able to simultaneously represent surface
  warming and the 0-700m and 700-2000m heat storage in a reasonable
  fashion.}
\label{4lyrdw}
\end{figure}

Figure.~\ref{4lyrdw} shows the response of such an extended four layer
model (Expt. 6 in Table.~1). Both SAT and OHC representations are
reasonable enough to suggest that this is a simple extension of the
two layer EBM that overcomes the shortcoming of the latter of not
being able to simultaneously represent the surface responses and
subsurface ocean heat uptake in a variety of settings.

\section{Discussion and Conclusions}
Simple climate models play a valuable role in helping interpret both
observations and responses of comprehensive earth system
models. To this end, we considered the anomaly-diffusion energy
balance model and used it in a Bayesian framework to interpret a few
CMIP5 experiments---the abrupt 4$\times$CO$_2$ forcing, the 1\% CO$_2$
forcing and the historical forcing experiments. The intention of the
CMIP abrupt 4$\times$CO$_2$ forcing experiment is one of calibration
and diagnosis of equilibrium climate sensitivity since post-industrial
surface warming itself is compatible with a wide range of climate
sensitivities. As such, we used the abrupt 4$\times$CO$_2$ forcing
CCSM4 experiment in conjunction with an observational estimate of OHC
to infer the SCM parameters and used those parameters to compute the
SCM responses to other forcings. We demonstrated that the
commonly-used anomaly-diffusing energy balance SCM is deficient in
being unable to simultaneously properly represent the SAT responses
and observational estimates of OHC from NODC. 

After verifying that this deficiency is not related to vertical
resolution, we considered the possibility that it is due to the
anomaly diffusion approximation. To this end, we demonstrated
mathematically that when outcropping isopycnals are considered in a
framework of horizontally-averaged layers, the outcropping of
isopycnals leads to non-local interactions in the framework of
horizontally-averaged layers. With this modification, we then showed
that four layers were sufficient to successfully resolve the
shortcoming of the AD-EBM class of SCMs.  We also go on to show that
that with such {\em apparent} non-local interactions---see supplement section
``Further Experiments with Non Local Interactions'' for a particular
form---a continuous version of the model is similarly able to
successfully resolve the shortcoming.

It is possible, however, that other kinds of augmentations to the
commonly-used AD-EBM may fix the identified deficiency as well. On the
other hand, since the simple extension that allows for simultaneously
properly representing the surface warming and the 0-700m and 700-2000m
heat storage is consistent with the effects of processes such as
subduction and mode water formation and deepwater formation on ocean
heat storage, our finding may be interpreted as indicative of the
importance of direct sequestration of heat in the vertical interior of
the world oceans by such processes. In particular, and as discussed in
the introduction, since the (deep) ocean loses heat to the atmosphere
at high latitudes through convective instabilities, when the stability
of the water column is incrementally enhanced in a warming scenario
(that is, the water column continues to remain unstable, but its
degree of instability is slightly reduced), the heat loss from the
deep ocean is slightly reduced leading effectively to a warming of the
deep. At the next order, the resulting slowdown of the overturning
circulation itself would further affect the heat uptake. Indeed, to
examine such effects, we note that in ongoing work we consider
extensions to a class of models that are used to study the overturning
circulation \citep{tziperman1986role, gnanadesikan1999simple,
  goodwin2012isopycnal, marshall2014conceptual} to simultaneously
represent surface warming and deeper ocean warming.

In both the four-layer and continuous variants of the EBM with
non-local interactions, we also find that being able to reasonably
represent the surface response and the 0-2000 m OHC is often
associated with heat storage at depths below 2000 m that is greater
than 10\%. Clearly this finding needs to be investigated further in a
hierarchical framework---a framework that will allow us to infer model
parameters using the multi-model CMIP5 ensemble. In ongoing work, we
are also examining (a) the implications of this study on estimates of
uncertainty of climate sensitivity and (b) if these modified SCMs will
help in better analyzing and inter-comparing ESMs.

\section*{Acknowledgements}
BTN would like to thank W. Riley Casper for help in identifying the
zero entries of $\widetilde {\vv A}$ in Eq.~10. This research was
supported by the U.S. Department of Energy (DOE) Office of Science
(Biological and Environmental Research), Early Career Research
program. All of the data used in this article has been previously
archived and may be obtained as follows: The Levitus ocean heat
content data may be obtained from the the National Oceanographic Data
Center at https://www.nodc.noaa.gov. CMIP5 data may be obtained from
one of the Earth System Grid Federation nodes, e.g.,
https://esgf-node.jpl.nasa.gov. The SAT data may be obtained from
https://crudata.uea.ac.uk/cru/data/temp-erature The historical
radiative forcing and the RCP8.5 scenario forcing may be obtained from
\citep{meinshausen2011rcp}. We acknowledge the World Climate Research
Programme's Working Group on Coupled Modelling, which is responsible
for CMIP, and we thank the climate modeling groups (listed in Table XX
of this paper) for producing and making available their model
output. For CMIP the U.S. Department of Energy's Program for Climate
Model Diagnosis and Intercomparison provides coordinating support and
led development of software infrastructure in partnership with the
Global Organization for Earth System Science Portals

\clearpage

\bibliographystyle{spbasic}
\bibliography{ebm}

\begin{thebibliography}{38}
\providecommand{\natexlab}[1]{#1}
\providecommand{\url}[1]{{#1}}
\providecommand{\urlprefix}{URL }
\expandafter\ifx\csname urlstyle\endcsname\relax
  \providecommand{\doi}[1]{DOI~\discretionary{}{}{}#1}\else
  \providecommand{\doi}{DOI~\discretionary{}{}{}\begingroup
  \urlstyle{rm}\Url}\fi
\providecommand{\eprint}[2][]{\url{#2}}

\bibitem[{Aldrin et~al(2012)Aldrin, Holden, Guttorp, Skeie, Myhre, and
  Berntsen}]{aldrin2012bayesian}
Aldrin M, Holden M, Guttorp P, Skeie RB, Myhre G, Berntsen TK (2012) Bayesian
  estimation of climate sensitivity based on a simple climate model fitted to
  observations of hemispheric temperatures and global ocean heat content.
  Environmetrics 23(3):253--271

\bibitem[{Balmaseda et~al(2013)Balmaseda, Trenberth, and
  K{\"a}ll{\'e}n}]{balmaseda2013distinctive}
Balmaseda MA, Trenberth KE, K{\"a}ll{\'e}n E (2013) Distinctive climate signals
  in reanalysis of global ocean heat content. Geophysical Research Letters
  40(9):1754--1759

\bibitem[{Bodman and Jones(2016)}]{bodman2016bayesian}
Bodman RW, Jones RN (2016) Bayesian estimation of climate sensitivity using
  observationally constrained simple climate models. Wiley Interdisciplinary
  Reviews: Climate Change 7(3):461--473

\bibitem[{Bryan and Lewis(1979)}]{bryan1979water}
Bryan K, Lewis L (1979) A water mass model of the world ocean. Journal of
  Geophysical Research: Oceans 84(C5):2503--2517

\bibitem[{Budyko(1969)}]{budyko1969effect}
Budyko MI (1969) The effect of solar radiation variations on the climate of the
  earth. Tellus 21(5):611--619

\bibitem[{Cheng et~al(2017)Cheng, Trenberth, Fasullo, Boyer, Abraham, and
  Zhu}]{cheng2017improved}
Cheng L, Trenberth KE, Fasullo J, Boyer T, Abraham J, Zhu J (2017) Improved
  estimates of ocean heat content from 1960 to 2015. Science Advances
  3(3):e1601,545

\bibitem[{Domingues et~al(2008)Domingues, Church, White, Gleckler, Wijffels,
  Barker, and Dunn}]{domingues2008improved}
Domingues CM, Church JA, White NJ, Gleckler PJ, Wijffels SE, Barker PM, Dunn JR
  (2008) Improved estimates of upper-ocean warming and multi-decadal sea-level
  rise. Nature 453(7198):1090--1093

\bibitem[{Friend(2011)}]{friend2011response}
Friend A (2011) Response of earth's surface temperature to radiative forcing
  over {AD} 1--2009. Journal of Geophysical Research: Atmospheres 116(D13)

\bibitem[{Geoffroy et~al(2012)Geoffroy, Saint-Martin, and
  Ribes}]{geoffroy2012quantifying}
Geoffroy O, Saint-Martin D, Ribes A (2012) Quantifying the sources of spread in
  climate change experiments. Geophysical Research Letters 39(24)

\bibitem[{Geoffroy et~al(2013)Geoffroy, Saint-Martin, Olivi{\'e}, Voldoire,
  Bellon, and Tyt{\'e}ca}]{geoffroy2013transient}
Geoffroy O, Saint-Martin D, Olivi{\'e} DJ, Voldoire A, Bellon G, Tyt{\'e}ca S
  (2013) Transient climate response in a two-layer energy-balance model. {Part
  I}: Analytical solution and parameter calibration using {CMIP5 AOGCM}
  experiments. Journal of Climate 26(6):1841--1857

\bibitem[{Gnanadesikan(1999)}]{gnanadesikan1999simple}
Gnanadesikan A (1999) A simple predictive model for the structure of the
  oceanic pycnocline. Science 283(5410):2077--2079

\bibitem[{Goodwin(2012)}]{goodwin2012isopycnal}
Goodwin P (2012) An isopycnal box model with predictive deep-ocean structure
  for biogeochemical cycling applications. Ocean Modelling 51:19--36

\bibitem[{Gregory(2000)}]{gregory2000vertical}
Gregory JM (2000) Vertical heat transports in the ocean and their effect on
  time-dependent climate change. Climate Dynamics 16(7):501--515

\bibitem[{Held et~al(2010)Held, Winton, Takahashi, Delworth, Zeng, and
  Vallis}]{held2010probing}
Held IM, Winton M, Takahashi K, Delworth T, Zeng F, Vallis GK (2010) Probing
  the fast and slow components of global warming by returning abruptly to
  preindustrial forcing. Journal of Climate 23(9):2418--2427

\bibitem[{Hoffert et~al(1980)Hoffert, Callegari, and Hsieh}]{hoffert1980role}
Hoffert MI, Callegari AJ, Hsieh CT (1980) The role of deep sea heat storage in
  the secular response to climatic forcing. Journal of Geophysical Research:
  Oceans 85(C11):6667--6679

\bibitem[{Houghton et~al(1997)Houghton, Meira~Filho, Griggs, and
  Maskell}]{houghton1997introduction}
Houghton JT, Meira~Filho LG, Griggs DJ, Maskell K (1997) An introduction to
  simple climate models used in the IPCC Second Assessment Report. WMO; UNEP

\bibitem[{Johansson et~al(2015)Johansson, {O'Neill}, Tebaldi, and
  H{\"a}ggstr{\"o}m}]{johansson2015equilibrium}
Johansson DJ, {O'Neill} BC, Tebaldi C, H{\"a}ggstr{\"o}m O (2015) Equilibrium
  climate sensitivity in light of observations over the warming hiatus. Nature
  Climate Change 5(5):449--453

\bibitem[{Knutti and Hegerl(2008)}]{knutti2008equilibrium}
Knutti R, Hegerl GC (2008) The equilibrium sensitivity of the earth's
  temperature to radiation changes. Nature Geoscience 1(11):735--743

\bibitem[{Levitus et~al(2012)Levitus, Antonov, Boyer, Baranova, Garcia,
  Locarnini, Mishonov, Reagan, Seidov, Yarosh et~al}]{levitus2012world}
Levitus S, Antonov JI, Boyer TP, Baranova OK, Garcia HE, Locarnini RA, Mishonov
  AV, Reagan J, Seidov D, Yarosh ES, et~al (2012) World ocean heat content and
  thermosteric sea level change (0--2000 m), 1955--2010. Geophysical Research
  Letters 39(10)

\bibitem[{Marshall and Zanna(2014)}]{marshall2014conceptual}
Marshall DP, Zanna L (2014) A conceptual model of ocean heat uptake under
  climate change. Journal of Climate 27(22):8444--8465

\bibitem[{Mauritsen et~al(2012)Mauritsen, Stevens, Roeckner, Crueger, Esch,
  Giorgetta, Haak, Jungclaus, Klocke, Matei et~al}]{mauritsen2012tuning}
Mauritsen T, Stevens B, Roeckner E, Crueger T, Esch M, Giorgetta M, Haak H,
  Jungclaus J, Klocke D, Matei D, et~al (2012) Tuning the climate of a global
  model. Journal of Advances in Modeling Earth Systems 4(3)

\bibitem[{Meinshausen et~al(2011{\natexlab{a}})Meinshausen, Raper, and
  Wigley}]{meinshausen2011emulating}
Meinshausen M, Raper SC, Wigley TM (2011{\natexlab{a}}) Emulating coupled
  atmosphere-ocean and carbon cycle models with a simpler model, {MAGICC6--Part
  1: Model} description and calibration. Atmospheric Chemistry and Physics
  11(4):1417--1456

\bibitem[{Meinshausen et~al(2011{\natexlab{b}})Meinshausen, Smith, Calvin,
  Daniel, Kainuma, Lamarque, Matsumoto, Montzka, Raper, Riahi
  et~al}]{meinshausen2011rcp}
Meinshausen M, Smith SJ, Calvin K, Daniel JS, Kainuma M, Lamarque J, Matsumoto
  K, Montzka S, Raper S, Riahi K, et~al (2011{\natexlab{b}}) The {RCP}
  greenhouse gas concentrations and their extensions from 1765 to 2300.
  Climatic change 109(1-2):213

\bibitem[{Murphy(1995)}]{murphy1995transient}
Murphy J (1995) Transient response of the {Hadley Centre} coupled
  ocean-atmosphere model to increasing carbon dioxide. {Part III}: analysis of
  global-mean response using simple models. Journal of Climate 8(3):496--514

\bibitem[{Nadiga and Urban(2016)}]{nadiga16}
Nadiga B, Urban N (2016) Dependence of inferred climate sensitivity on the
  discrepancy model, climate informatics. Proceedings of the 6th International
  Workshop on Climate Informatics: CI 2016 NCAR Technical Notes
  NCAR/TN-529+PROC, doi: 10.5065/D6K072N6:29--32

\bibitem[{North et~al(1981)North, Cahalan, and Coakley}]{north1981energy}
North GR, Cahalan RF, Coakley JA (1981) Energy balance climate models. Reviews
  of Geophysics 19(1):91--121

\bibitem[{Padilla et~al(2011)Padilla, Vallis, and
  Rowley}]{padilla2011probabilistic}
Padilla LE, Vallis GK, Rowley CW (2011) Probabilistic estimates of transient
  climate sensitivity subject to uncertainty in forcing and natural
  variability. Journal of Climate 24(21):5521--5537

\bibitem[{Pedlosky(2013)}]{pedlosky2013ocean}
Pedlosky J (2013) Ocean circulation theory. Springer Science \& Business Media

\bibitem[{Raper and Cubasch(1996)}]{raper1996emulation}
Raper S, Cubasch U (1996) Emulation of the results from a coupled general
  circulation model using a simple climate model. Geophysical Research Letters
  23(10):1107--1110

\bibitem[{Raper et~al(2001)Raper, Gregory, and Osborn}]{raper2001use}
Raper S, Gregory JM, Osborn T (2001) Use of an upwelling-diffusion energy
  balance climate model to simulate and diagnose {A/OGCM} results. Climate
  Dynamics 17(8):601--613

\bibitem[{Raper et~al(2002)Raper, Gregory, and Stouffer}]{raper2002role}
Raper SC, Gregory JM, Stouffer RJ (2002) The role of climate sensitivity and
  ocean heat uptake on {AOGCM} transient temperature response. Journal of
  Climate 15(1):124--130

\bibitem[{Schneider and Thompson(1981)}]{schneider1981atmospheric}
Schneider SH, Thompson SL (1981) Atmospheric {CO2} and climate: importance of
  the transient response. Journal of Geophysical Research: Oceans
  86(C4):3135--3147

\bibitem[{Sellers(1969)}]{sellers1969global}
Sellers WD (1969) A global climatic model based on the energy balance of the
  earth-atmosphere system. Journal of Applied Meteorology 8(3):392--400

\bibitem[{Talley(2011)}]{talley2011descriptive}
Talley LD (2011) Descriptive physical oceanography: an introduction. Academic
  press

\bibitem[{Taylor et~al(2012)Taylor, Stouffer, and Meehl}]{taylor2012overview}
Taylor KE, Stouffer RJ, Meehl GA (2012) An overview of cmip5 and the experiment
  design. Bulletin of the American Meteorological Society 93(4):485--498

\bibitem[{Tziperman(1986)}]{tziperman1986role}
Tziperman E (1986) On the role of interior mixing and air-sea fluxes in
  determining the stratification and circulation of the oceans. Journal of
  Physical Oceanography 16(4):680--693

\bibitem[{Urban and Keller(2010)}]{urban2010probabilistic}
Urban NM, Keller K (2010) Probabilistic hindcasts and projections of the
  coupled climate, carbon cycle and atlantic meridional overturning circulation
  system: A {Bayesian} fusion of century-scale observations with a simple
  model. Tellus A 62(5):737--750

\bibitem[{Wigley and Raper(1992)}]{wigley1992lmplications}
Wigley T, Raper S (1992) lmplications for climate and sea level of revised
  {IPCC} emission scenarios. Nature 357:28

\end{thebibliography}

\clearpage
\appendix
\section*{Supplementary Material}

\title{Improved Representation of Ocean Heat Content in Energy Balance
  Models. Supplementary Material}
\titlerunning{Supplementary Material for Representation of OHC in EBMs}
\author{B.T. Nadiga and N.M. Urban}
\institute{B.T. Nadiga \at
Los Alamos National Lab, Los Alamos, NM 87544\\
Tel.: 505-667-9466\\
\email{balu@lanl.gov}
\and
N.M. Urban \at
Los Alamos National Lab, Los Alamos, NM 87544
}
\date{}
\maketitle
\section{Further Methodological Details}\label{appa}
EBMs have been used extensively to obtain estimates of Equilibrium
Climate Sensitivity (ECS) of AOGCMs and ESMs. However, unlike the
comprehensive AOGCMs and ESMs, the simpler of the EBMs such as the
ones considered in this article do not represent natural climate
variability. For this reason, the Bayesian calibration procedure
requires the specification of a structure for natural variability.
For example, for the temperature of the upper layer, $T_u$,
\begin{linenomath*}
\begin{align}
T_u^{ESM}(t) = T_u^{EBM}(t) + \epsilon_t, 
\end{align}
\end{linenomath*}
where $\epsilon_t$ is the discrepancy and similarly for other
variables such as radiative imbalance and ocean heat
content. Depending on the variable, the discrepancy structure can be
reasonably approximated by simple stochastic processes such as
independent and identically distributed (iid),
auto-regressive processes (AR; e.g., AR(1)) or Gaussian Processes
(GP). As mentioned in the main body, we used AR(1) for each of the
variables considered in the Bayesian calibration procedure.

There is, however, a dependency of parameter estimates and their
uncertainties on the assumed temporal structure of the
discrepancy. Likewise, how the posterior distribution of calibrated
responses match the original data (observational estimates or ESM
responses) depends on the discrepancy model used.
See \cite{nadiga16} for an example of how ECS
estimates and their uncertainty depends on the structure assumed for
the discrepancy in the context of the two-layer EBM considered here.

Even after the structure of the discrepancy model is assumed, the
results of the Bayesian inference procedure depend on whether the
parameters of the discrepancy model are specified or inferred. A
number of the experiments presented here were performed both ways:
once while inferring, also, the parameters of the AR(1) model and a
second time while specifying the parameters of the AR(1) model after
first fitting that particular response alone using least-squares
(i.e., after obtaining a point estimate of only the EBM
parameters). In most of the cases, we found that attempting to infer
the parameters of the AR(1) model simultaneously led to issues of
identifiability. That is, for a significant number of the parameters
related to the discrepancy model that we were, also, trying to infer,
the calibration data were unable to move the prior distribution of
that parameter. That is, for these parameters, the posterior
distribution remained very close to the specified prior
distribution. Consequently, the width of the assumed uninformative
priors for these parameters led to further augmenting uncertainties in
the EBM parameters. For this reason, it was deemed more appropriate to
perform the inferences while specifying the parameters for the
discrepancy model.

The covariance structure for the AR(1) process can be written as 
\begin{linenomath*}
\begin{align}
\Sigma (t-s) =   \sigma^2 \rho^{|t-s|}
\end{align}
\end{linenomath*}
The specified values of $\sigma$ and $\rho$ were as follows abrupt4x
SAT: (0.15 K, 0.5); abrupt4x RN: (0.27 W/m$^2$, 0.1); 0-700 m OHC: (2
$\times 10^{22}$ J, 0.8); 700-2000 m OHC: (1 $\times 10^{22}$ J, 0.8).
This is consistent with the anamolous
forcing varying fastest, the SAT response which is an integrated
response to the anamolous forcing varying on a slower time scale
related to the lower heat capacity of the upper layer and the OHC
varying slowest because of its higher heat capacity.
There may be legitimate reasons to consider other values for these
parameters. Further experimentation was conducted along these
lines. Results of such experimentation suggest that there may be a
tradeoff between fitting the various pieces of calibration data but
that such tradeoff will not affect the overall nature of our results and
conclusions.

\renewcommand{\fgprfx}{CCSM4-SAT-RAD-2-clbr2}
\begin{figure}
\includegraphics[width=0.5\textwidth]{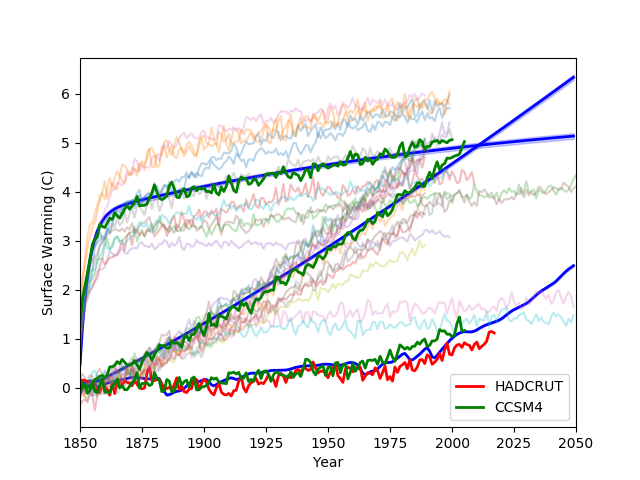}
\renewcommand{\fgprfx}{2-clbr0}
\includegraphics[width=0.5\textwidth]{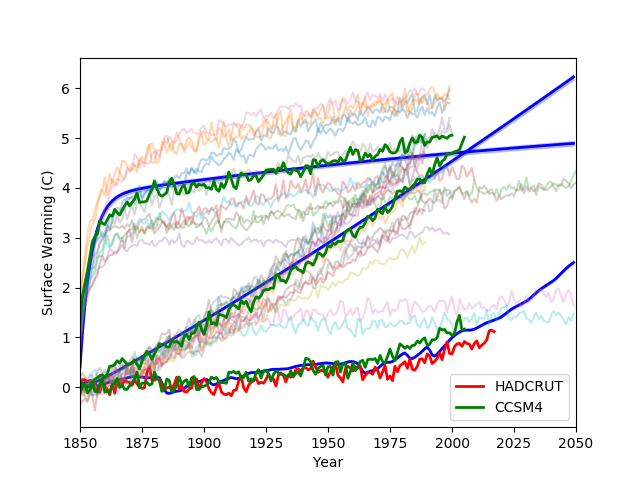}
\caption{SAT responses of the two-layer EBM in experiments 2 and 3.}
\label{sat2}
\end{figure}

\begin{figure}
\renewcommand{\fgprfx}{CCSM4-SAT-2-clbr1}
\includegraphics[width=0.5\textwidth]{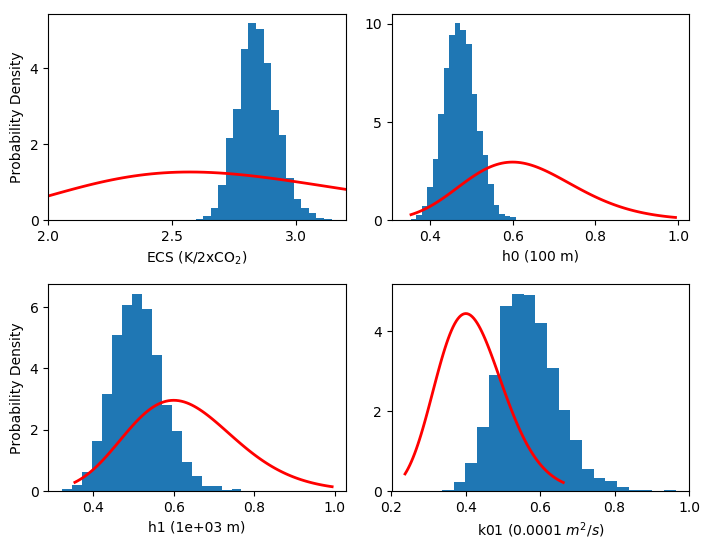}
\renewcommand{\fgprfx}{CCSM4-SAT-RAD-2-clbr2}
\includegraphics[width=0.5\textwidth]{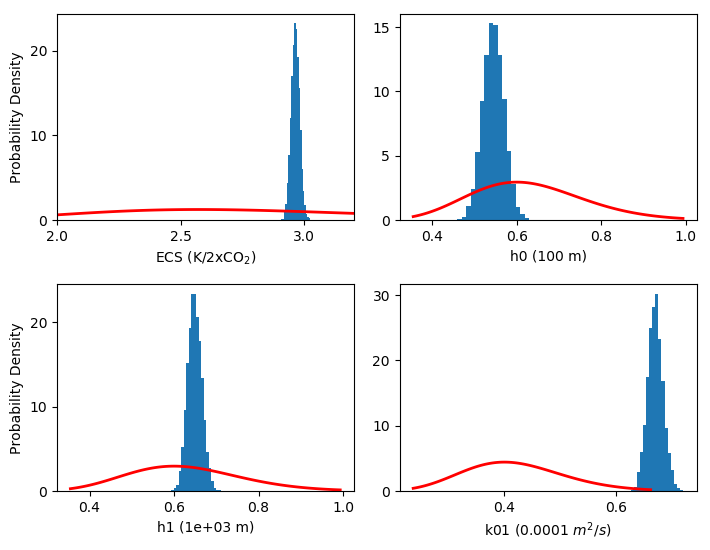}
\caption{Prior (red curves) and posterior (blue histograms) distribution of the four model parameters
  in the two layer EBM in experiments 1 (left) \& 2
  (right). Constraining the overall heat storage in experiment 2 leads
to slight changes in mode of the inferred parameters, but to large
reductions in the width of the posteriors.}
\label{pdfs-12}
\end{figure}

\begin{figure}
\renewcommand{\fgprfx}{CCSM4-SAT-2-clbr1}
\includegraphics[width=0.5\textwidth]{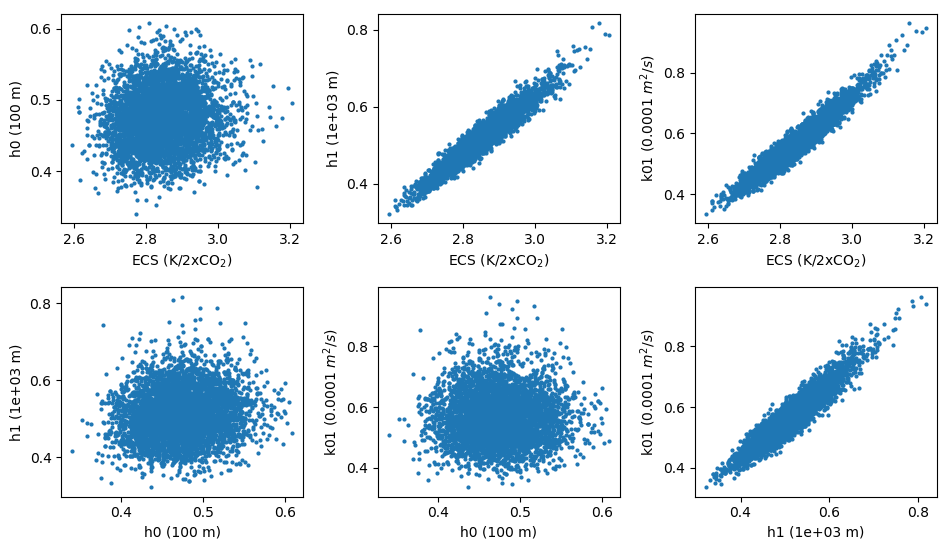}
\renewcommand{\fgprfx}{CCSM4-SAT-RAD-2-clbr2}
\includegraphics[width=0.5\textwidth]{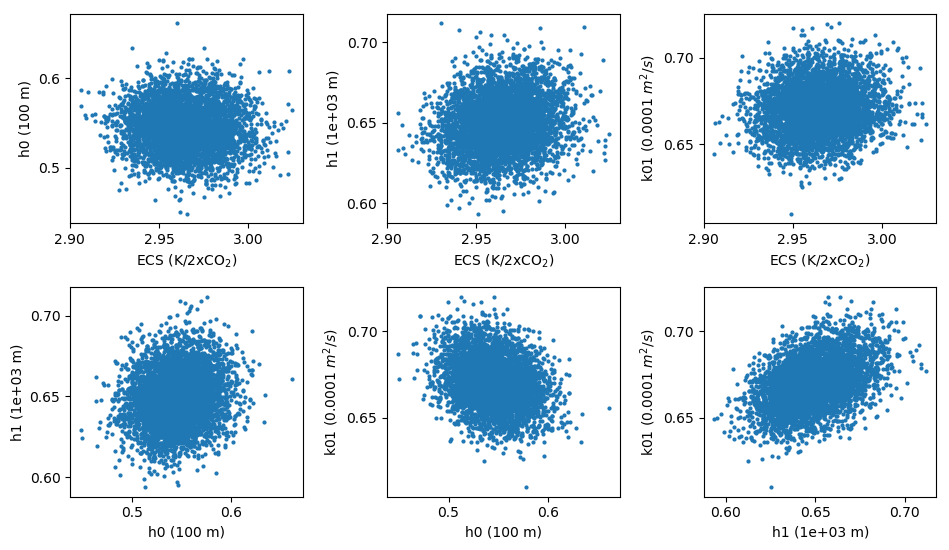}
\caption{Posterior parameter correlations in experiments 1 \&
  2. Constraining the overall heat storage in experiment 2 leads to
  removal of the correlations seen in experiment 1 to a large extent.}
\label{pcorr-12}
\end{figure}

\renewcommand{\fgprfx}{2-clbr0}
\begin{figure}
\includegraphics[width=0.385\textwidth]{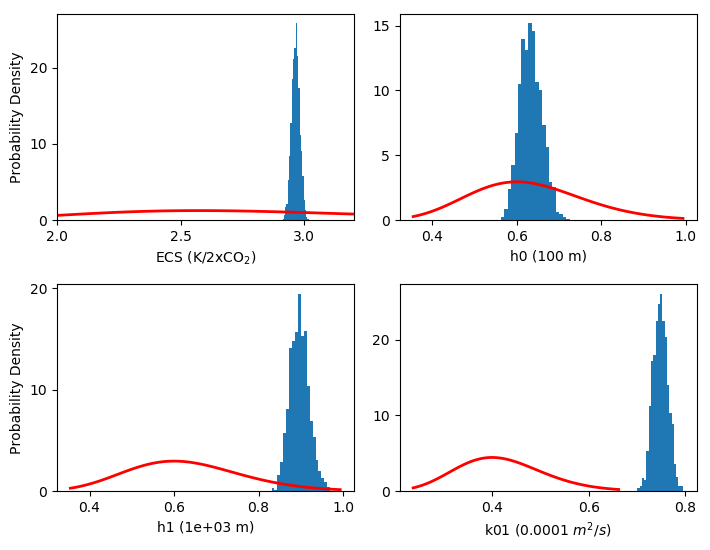}
\includegraphics[width=0.52\textwidth]{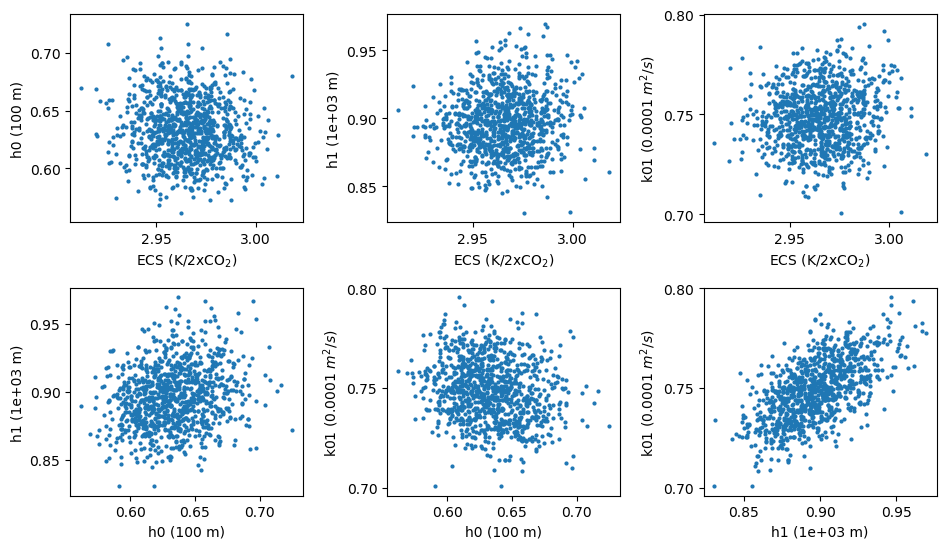}
\caption{Parameter distributions, and correlations in
  experiment 3. Constrainting 0-700 m and 700-2000 m OHC leads to
  further changes in the inferred two-layer EBM parameters, but as
  discussed in Sec. 2, the two-layer EBM is unable to reasonably fit
  the observational estimates of OHC.}
\label{expt3}
\end{figure}

\section{Further Two-Layer EBM Details}\label{appb}
As mentioned in Sec.~2 of the article and seen in the top row of
Fig.~2 of the article, while there
are differences in representation of OHC in the two-layer EBM in
experiments 1 and 2, these differences are not dramatic. For
completeness, the SAT response is shown in the left panel of
Fig.~\ref{sat2} and the differences in the SAT responses are seen to
be minor as well.

However, constraining the total heat storage in the climate
system in experiment 2 by calibrating against the radiative
imbalance/nonequilibrium (RI/RN)
response of the ESM results in significant changes to the inferred
(posterior) distributions of the SCM parameters. As seen in
Fig.~\ref{pdfs-12}, using the RN constraint leads to the  inferred EBM
parameters being significantly different. It is also seen in that
figure that the posterior distributions themselves are much narrower.

Furthermore, correlations that exist between
model parameters in experiment 1 are seen to be greatly reduced in
experiment 2 (see Fig.~\ref{pcorr-12}).

For completeness, SAT responses in experiment 3 is shown in the right
panel of Fig.~1 and the parameter distributions, and their
correlations in experiment 3 are shown in Fig.~\ref{expt3}.

\begin{table}
\centering
{\tabulinesep=1.2mm

\begin{tabu}{|c|c|c|}
\hline
\multirow{2}{*}{Expt. \#} & \multirow{2}{*}{SCM}                          & \multirow{2}{*}{Comment}            \\
                          &                                               &                                     \\
\hhline{|=|=|=|}
\multirow{3}{*}{5a}       & 40L; MLD inferred                             &                                     \\
                          & $\kappa_v^{TC}$ inferred                      & Degraded 4x SAT                     \\
                          & $\kappa_v^{ML_{bot}} = \kappa_v^{TC}$         & Poor OHC                            \\
\hline
\multirow{3}{*}{5b}       & \multirow{2}{*}{40L}                          & \multirow{2}{*}{Degraded 4x SAT}    \\
                          &                                               &                                     \\
                          & MLD = 70 m                                    & Poor OHC                            \\
\hline
\multirow{3}{*}{5c}       & 40L                                           & \multirow{2}{*}{Degraded 4x SAT}    \\
                          & $\kappa_v^{ML_{bot}}\ne\kappa_v^{TC}$         &                                     \\
                          & $\kappa_v^{ML_{bot}}, \kappa_v^{TC}$ inferred & Poor OHC                            \\
\hline
\multirow{3}{*}{5d}       & 40L                                           & Poor SAT                            \\
                          & $\kappa_v^{TC} = 10^{-5}$ m$^2$/s             & Poor RN                             \\
                          & $\kappa_v^{ML_{bot}}\ne\kappa_v^{TC}$         & Poor OHC                            \\
\hline
\multirow{3}{*}{5e}       & 40L; w inferred                               & Same as in Expt. 5c                 \\
                          & $\kappa_v^{ML_{bot}}\ne\kappa_v^{TC}$         & Degraded 4x SAT                     \\
                          & $\kappa_v^{ML_{bot}}, \kappa_v^{TC}$ inferred & Poor OHC                            \\
\hline
\multirow{3}{*}{5f}       & 40L; $\kappa_v^{>650 m}$ inferred             & Same as in Expt. 5c                 \\
                          & $\kappa_v^{ML_{bot}}\ne\kappa_v^{TC}$         & Degraded 4x SAT                     \\
                          & $\kappa_v^{ML_{bot}}, \kappa_v^{TC}$ inferred & Poor OHC                            \\
\hline
5g                        & 60L; Same as 5c                               & Same as in Expt. 5c                 \\
\hline
\multirow{2}{*}{7a}        & 40L; Same as 5c                               & Reasonable SAT                      \\
                          & Non-local interaction                         
                          & Reasonable OHC                                                                      \\
\hline
\multirow{2}{*}{7b}       & 60L; Same as 5c/g/7                               & \multirow{2}{*}{Same as in Expt. 7a} \\
                          & Non-local interaction                         &                                     \\
\hline
\end{tabu}}
\caption{Some of the continuous AD-EBM variants considered}
\label{tab2}
\end{table}

\begin{table}
\centering
{\tabulinesep=1.2mm

  \begin{tabu}{|c|c|c|c|c|c|c|}
    \hline
    \multirow{2}{*}{Expt.}&\multicolumn{4}{c|}{SAT RMS
      Error}&\multicolumn{2}{c|}{OHC RMS Error}\\
    \cline{2-7}
    &4x&1\%&Hi&HC&0-700m&700-2000m\\
\hhline{|=|=|=|=|=|=|=|}
5a&0.19&0.11&0.16&0.17&7.5&2.2\\
\hline
5b&0.26&0.12&0.16&0.16&8.2&2.3\\
\hline
5c&0.18&0.13&0.16&0.18&6.5&1.7\\
\hline
5d&0.39&0.14&0.15&0.17&7.0&3.1\\
\hline
5e&0.16&0.13&0.16&0.18&5.9&1.4\\
\hline
5f&0.17&0.13&0.16&0.18&6.5&1.7\\
\hline
5g&0.18&0.13&0.16&0.18&6.5&1.7\\
\hline
7a&0.15&0.12&0.16&0.18&2.0&1.5\\
\hline
7b&0.15&0.12&0.16&0.18&1.6&1.5\\
\hline
\end{tabu}}
\caption{RMS error of SAT, with respect to CCSM4 experiments abrupt4x,
  1\%, historical forcing, and with respect to HadCRUT, and RMS error
  of OHC with respect to 0-700m and 700-2000m estimates from NODC in
  historical-forcing runs. Units for RMS error of SAT is degrees C,
  and units for RMS error of OHC is $10^{22}$ J.}
\label{tab2}
\end{table}

\renewcommand{\fgprfx}{../SantaFe/A}
\begin{figure}
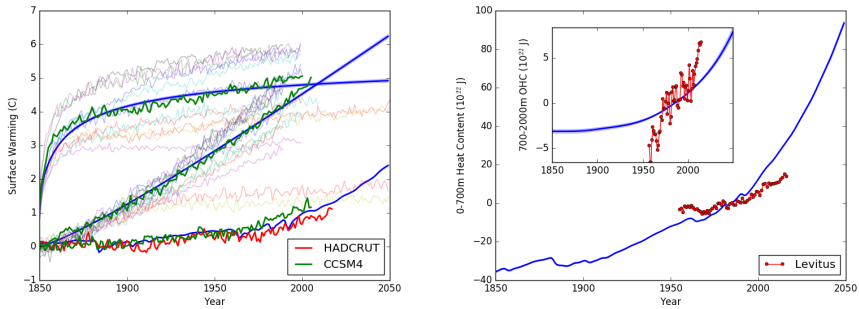

\satohc{40-clbr0-same_mlbotdff}
\caption{In Experiment 5a, allowing 40 layers in the AD-EBM model does
  not help improve the reprsentation of OHC.}
\label{40lyr}
\end{figure}

\begin{figure}
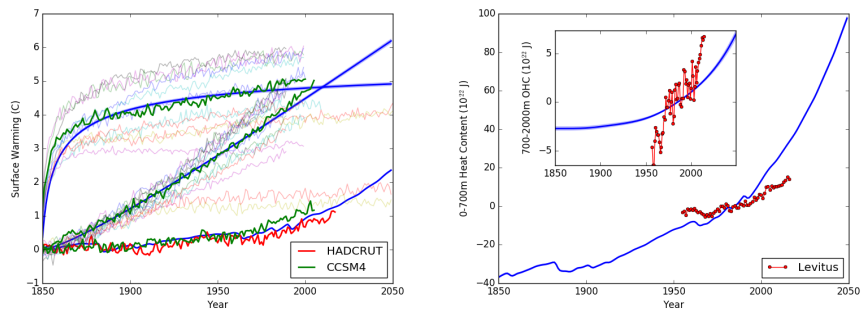

\satohc{40-clbr0-fxdh0-same_mlbotdff}
\caption{In experiment 5b, a variant of experiment 5a, the upper layer
  depth is held fixed at 70 m. This does not help improve the OHC
  response and the results are similar to that of experiment 5a.}
\label{40fxdh0}
\end{figure}

\begin{figure}
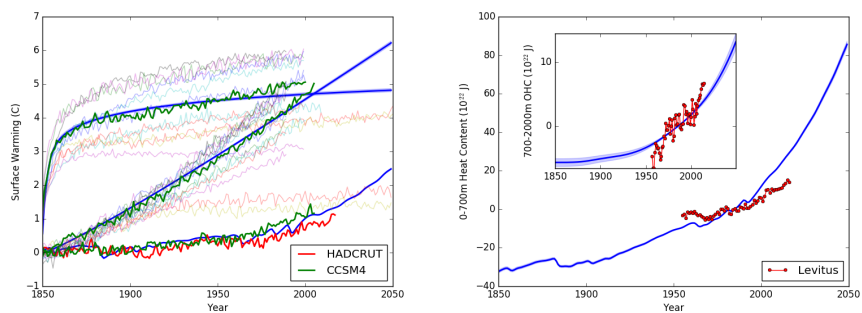

\satohc{40}
\caption{In Experiment 5c, it is seen that allowing vertical
  diffusivity at the bottom of the ML to be different from its
  thermocline value leads to improvement in the SCM response. However,
  the improved response is comparable to that of the two-layer EBM and
  suffers from similar shortcomings.}
\label{40k01k12}
\end{figure}

\begin{figure}
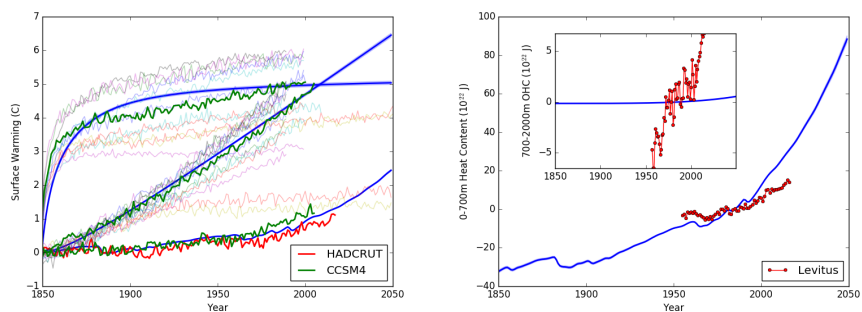

\satohc{40-clbr0-1e-5}
\caption{In Experiment 5d, a variant of experiment 5c, the thermocline
  diffusivity is set at a realistic value of 10$^{-5}$ m$^2$/s. While
  not helping improve the representation of OHC, the SAT response in
  the abrupt4x case is seen to be degraded.}
\label{5d}
\end{figure}

\begin{figure}
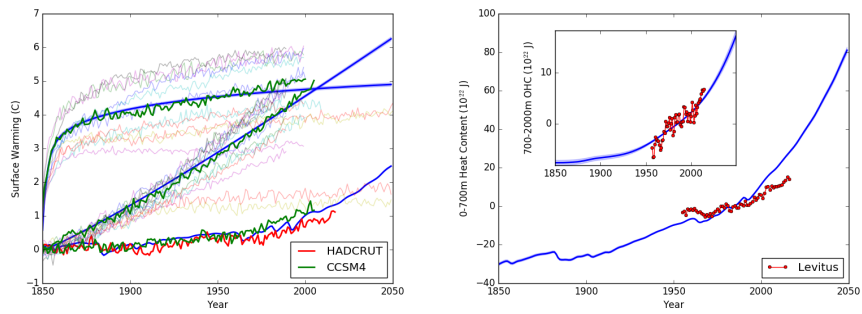

\satohc{40-clbr0-w}
\caption{In Experiment 5e, a variant of experiment 5c, a
  depth-independent upwelling velocity is allowed.  Allowing upwelling
  does not seem to help improve the representation of OHC.}
\label{wvel}
\end{figure}

\begin{figure}
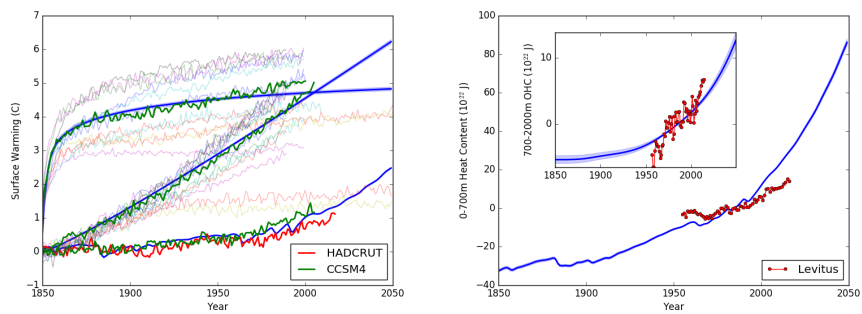

\satohc{40-clbr0-kv700}
\caption{In Experiment 5f, a variant of experiment 5c, the vertical
  diffusivity below 650 m is allowed to take on a different inferred
  value. Again the differences in responses from 5c are
seen to be minor.}
\label{kv700}
\end{figure}

\begin{figure}
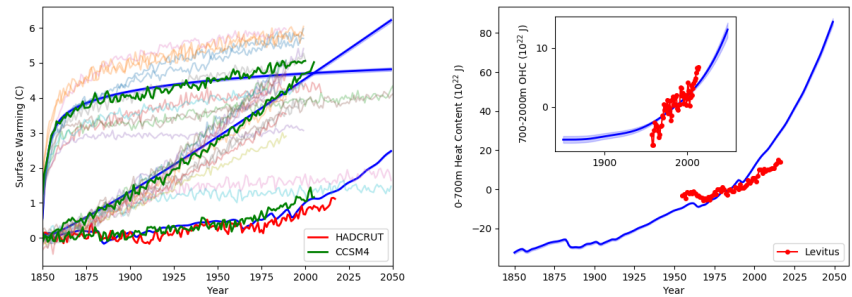

\satohc{60-clbr0}
\caption{In Experiment 5g, a variant of experiment 5c, the number of
layers is increased to 60. The differences in responses from 5c are
seen to be very minor.}
\label{60lyr}
\end{figure}

\section{Experiments with Continuous  AD-EBM Variants}\label{appc}
Table.~\ref{tab2} lists some of the continuous variants of AD-EBM
discussed here. We note that the calibration procedure in each of
these experiments were identical to that in Experiment 3 (see Table
1). In the first version (Experiment 5a), we consider a
variable thickness top layer and discretize the sub-surface
thermocline region down to 2500 m into 39 equi-depth
layers. Figure~\ref{40lyr} shows that the continuously-stratified
AD-EBM suffers from a similar shortcoming in not being able to
simultaneously properly represent surface warming and the 0-700m and
700-2000m heat storage.

In Experiment 5a, the thickness of the top layer in the continuously
stratified AD-EBM was seen to be very shallow ($\sim$ 10 m).
For this reason, we conduct Experiment 5b in which the thickness of
the top layer is fixed at 70 m. The SAT and OHC behavior is shown in
Fig.~\ref{40fxdh0}. Fixing the top layer depth at a reasonable
value seem to have little effect on the representation of heat
storage, and the results are similar to that in Experiment 5a.

In Experiment 5c, the vertical diffusivity at the bottom of the top
layer is allowed to be different from its thermocline value in order to see if
this will permit a better representation of ocean heat content. Significant
improvements are indeed seen in the responses of the SCM and
qualitatively, these responses are similar to those in the two-layer
model. Indeed, the results of this experiment seem to strongly suggest
that the continuously-stratified AD-EBM suffers from a similar
shortcoming in not being able to simultaneously properly represent
surface warming and the 0-700m and 700-2000m heat storage.

Next, we consider vertical diffusivity to have realistic values of
10$^{-5}$ m$^2$/s in the thermocline and 10$^{-4}$ m$^2$/s in the
abyssal ocean with a vertical profile following Bryan and Lewis,
1979\nocite{bryan1979water}. However, the transition from the
thermocline value to the abyssal value occurs (and is conventionally
thought to be so) at 2500 m. As discussed in the main body,
observational estimates of present day heating below 2000 m is small
and so, in Experiment 5d, the value of vertical diffusivity is fixed
at the thermocline value of 10$^{-5}$ m$^2$/s. 
Experiment 5d was among the worst in 
being able to match the ESM responses and the observed estimates of OHC.

Experiment 5e differs from Experiment 5c in allowing for a depth-independent 
upwelling velocity that is inferred in the Bayesian calibration procedure. 
The degree of mismatch between the SCM responses and the ESM responses
and observational estimates of OHC are similar to that in Experiment 5c.

In experiment 5f, a variant of 5c, we allow the diffusivity to take on
a different value below 650 m. This was motivated more by the nature
of the observational estimate of OHC rather than physics. However,
only minor differences are seen with respect to experiment 5c.
 
In Experiment 5g, we increase the number of layers in the thermocline
to 60, and Fig.~\ref{60lyr}, demonstrates convergent behavior of the
inability to simultaneously properly represent surface warming and the
0-700m and 700-2000m heat storage in the continuously stratified AD-
EBM.

The inability of the continuously stratified AD-EBM may be understood
as follows: Experiment 5c (equivalently 5g) fits ESM responses and
observational estimates of OHC best.  The inferred value of
thermocline diffusivity in these cases range between 0.7 and 0.9
cm$^2$/s. These values are much higher than the accepted levels of 0.1
cm$^2$/s. One reason for these large values of vertical diffusivity is
that the model is attempting to get sufficient heat down to the
700-2000 m depths in order to satisfy the observational
constraint. And in trying to do so, it is clear from these experiments
that this is causing too large a heating in the 0-700 m depths---a
feature that is consistent across a number of the experiments
described here.  That is, assuming that vertical diffusivity is not
varying with time (and there is no reason to expect it to have varied
substantially with time in the post industrial period) getting
observed levels of warming into the 700-2000 m layer while limiting
the warming of the 0-700m layer to observed levels is difficult to
achieve in the AD-EBM.

\section{Further Experiments with Non Local Interactions}\label{appd}



\begin{figure}
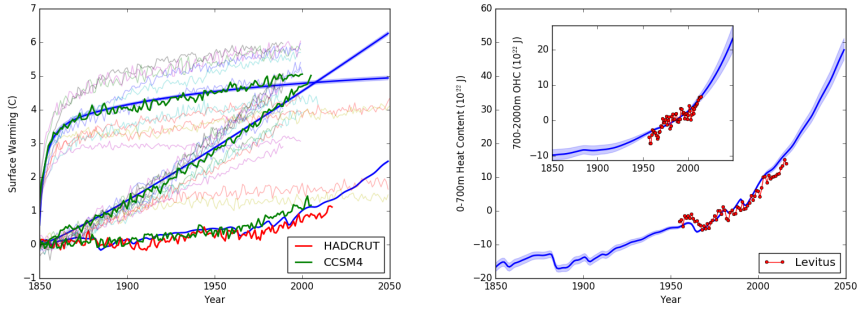

\satohc{40-clbr0-dw-1}
\caption{In Experiment 7a, a modification of the
  continuously-stratified model to allow for a non-local
  interaction---motivated by the direct sequestration of heat by
  deepwater formation processes---is seen to be able to simultaneously
  represent surface warming and the 0-700m and 700-2000m heat storage
  in a reasonable fashion.}
\label{40lyrdw}
\end{figure}

\begin{figure}
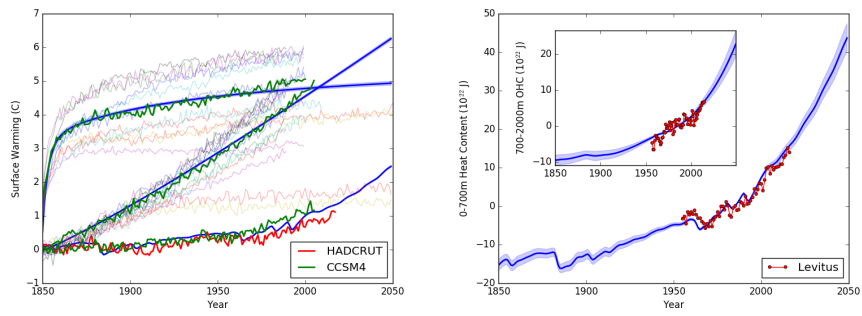

\satohc{60dw1}
\caption{In Experiment 7b, a variant of experiment 7, that has 60
  layers in the vertical, a convergent behavior with respect to
  increasing vertical resolution is seen.}
\label{60dw1}
\end{figure}

We next consider a modification of the continuous version of the
AD-EBM that addresses the deficiency. In the original approach of
Hoffert, 1980 to the AD-EBM, the {\em direct} sequestration of heat by
the polar downwelling branch in the cartoon of the overturning
circulation in the world oceans is not considered; only the effect of
the {\em resultant} weak and broad upwelling is considered. We
hypothesize that the deficiency of the AD-EBM identified above can be
remedied by introducing a parameterization of the {\em direct}
sequestration of heat by processes related to polar downwelling in the
AD-EBM. To test this hypothesis, we consider a parameterization of the
{\em direct} sequestration of heat by polar downwelling, and in
anology with the formation of the North Atlantic Deep Waters. The
structure of a nominal overturning cell in which the NADW is a
component has a maximum in the overturning streamfunction around a km
in depth. Consequently, we consider a profile of exchange of heat
between the first sub-surface layer and the interior that has a
Gaussian profile in the vertical centered at $z_{mx}$ and a vertical
scale of $\Delta z$, and where both $z_{mx}$ and $\Delta z$ are
inferred. That is, the anomaly-diffusion nature of the thermocline is
augmented as
\begin{linenomath*}
\begin{equation}
\frac{\partial T}{\partial t} + w \frac{\partial T}{\partial z} =
\frac{\partial}{\partial z} \left(\kappa(z)  \frac{\partial
    T}{\partial z} \right)
+ \gamma_d \exp\left(-\frac{(z -
      z_{mx})^2}{\Delta z^2}\right) (T_{1} - T)\notag.
\label{eqdw}
\end{equation}
\end{linenomath*}
Here $\gamma_d$ is a scalar coefficient of heat exchange associated
with the parameterization and is inferred in the calibration process.
Figure~\ref{40lyrdw} shows that the SAT and OHC responses are fit
reasonably in the model. Furthermore, in this simplified
parameterization, if half of the depth $z_{mx}$ is taken as the depth
at which the maximum of the overturning streamfunction occurs, the
maximum overturning depth is inferred to lie at about a km, which is
somewhat realistic. Finally, in Fig.~\ref{60dw1}, it is seen that the
changes are minimal when the number of vertical layers is increased
from 40 to 60.

\end{document}